\documentclass{elsart}
\usepackage{ifpdf}

\usepackage{graphicx}

\usepackage{amssymb}

\begin{document}

\begin{frontmatter}



\title{Performance of a fine-sampling electromagnetic calorimeter
  prototype in the energy range from 1 to 19 GeV}


\author{Yu.V.Kharlov\corauthref{cor1}},
\author{P.A.Semenov},
\author{Yu.A.Matulenko},
\author{O.P.Yushchenko},
\author{Yu.I.Arestov},
\author{G.I.Britvich},
\author{S.K.Chernichenko},
\author{Yu.M.Goncharenko},
\author{A.M.Davidenko},
\author{A.A.Derevschikov},
\author{A.S.Konstantinov},
\author{V.A.Kormilitsyn},
\author{V.I.Kravtsov},
\author{Yu.M.Melnik},
\author{A.P.Meschanin},
\author{N.G.Minaev},
\author{V.V.Mochalov},
\author{D.A.Morozov},
\author{A.V.Ryazantsev},
\author{I.V.Shein},
\author{A.P.Soldatov},
\author{L.F.Soloviev},
\author{A.V.Soukhikh},
\author{A.N.Vasiliev},
\author{M.N.Ukhanov},
\author{V.G.Vasilchenko\thanksref{deceased}},
\author{A.E.Yakutin}
\corauth[cor1]{corresponding author: \tt{Yuri.Kharlov@ihep.ru}}
\thanks[deceased]{deceased}

\address{Institute for High Energy Physics, Protvino, Russia}

\begin{abstract}
  The fine-sampling electromagnetic calorimeter prototype has been
  experimentally tested using the $1-19$~GeV/$c$ tagged beams of
  negatively charged particles at the U70 accelerator at IHEP,
  Protvino. The energy resolution measured by electrons is
  $\Delta{E}/E=2.8\%/\sqrt{E} \oplus 1.3\%$. The position resolution
  for electrons is $\Delta{x}=3.1 \oplus 15.4/\sqrt{E}$~mm in the
  center of the cell. The lateral non-uniformity of the prototype
  energy response to electrons and MIPs has turned out to be
  negligible. Obtained experimental results are in a good agreement
  with Monte-Carlo simulations.
\end{abstract}

\end{frontmatter}

\section*{Introduction}
\label{sec:Introduction}

Electromagnetic calorimeters are based on the total energy deposition
of photons and electrons in the active medium of detectors. Energy
deposited by secondary particles of an electromagnetic shower is
detected either as a Cherenkov radiation of electrons and positrons,
like in the lead-glass calorimeters \cite{lead-glass}, or as a
scintillation light emitted by the active medium
\cite{sci-calo}. Sampling calorimeters constructed from alternating
layers of organic scintillator and heavy absorber, have been used in
high energy physics over last tens of years \cite{lead-scin}. The
sampling of such calorimeters is determined by the required lateral
size of electromagnetic shower, expressed by a Moli\`ere radius $R_M$,
and the light yield provided by scintillator plates. Scintillation
light is absorbed, re-emitted and transported to a photodetector by
wave-length shifting (WLS) optical fibers penetrating through the
calorimeter modules longitudinally (along the beam direction). Typical
stochastic term of the energy resolution of all large electromagnetic
calorimeters of the sampling type was about 10\% \cite{PHENIX, HERA-B,
LHCB}.

Recently the improved electromagnetic calorimeter modules with a very
fine sampling have been developed for KOPIO experiment at BNL
\cite{KOPIO}. The energy resolution of these modules, measured with
photons of energy $220-350$~MeV, was about $3\%/\sqrt{E~\mbox{(GeV)}}$
\cite{Atoian2004}.  Details of the improved modules tested in the
energy range of $50-1000$~MeV, including mechanical construction,
selection of WLS fibers and photodetectors as well as development of a
new scintillator with improved optical and mechanical properties are
described in \cite{Atoian2008}.

Similar high-performance electromagnetic calorimeters are now being
considered for PANDA and CBM experiments
\cite{CBM-TSR,PANDA-TSR}\ at the future FAIR facility, which is under
construction at GSI, Darmstadt in Germany. The both fixed target
experimental setups require an ability to measure single photons,
$\pi^0$'s as well as $\eta$'s in the wide energy range with
excellent energy and position resolutions.  Fine-sampling 
calorimeters, not very expensive and meanwhile covering wide areas like
3~m$^2$ in PANDA and 100~m$^2$ in CBM, were chosen to meet the
requirements. The energy range in PANDA and CBM experiments will be
extended up to 15 and 35~GeV, respectively.  It is essential for these
projects to study parameters of a fine-sampling calorimeter in the
wide energy region, significantly wider than the one with the
existing data (only up to 1~GeV).

In this paper, we describe a fine-sampling electromagnetic calorimeter
prototype with lead absorber plates, which thickness is significantly
smaller than radiation length $X_0$ of lead. Such a small thickness of
the absorber layers results in a small interaction probability of the
secondary shower particles.  The design of this prototype is close to
the KOPIO one including the same lateral size of cells. The results of
miscellaneous studies of the prototype in the energy range from 1 to
19~GeV are presented in this paper.


\section{Design of the modules}
\label{sec:Design}

The electromagnetic calorimeter modules with fine sampling were
constructed at IHEP, Protvino. A module design was based on the
electromagnetic calorimeter for the KOPIO experiment, with additional
modification to high-energy range. Details of the mechanical
design of the modules can be found in \cite{Atoian2008}, but a few
modifications were applied to the prototype under study. The KOPIO
experiment was aimed to low-energy photons, and the total radiation
length $16X_0$ was enough for their purposes. The current prototype is
being proposed for CBM and PANDA, where the photon energy
extends up to 30~GeV, and the requirement to provide the total
radiation length of $20X_0$ was put to the design. The modules were
assembled from 380 alternating layers of lead and scintillator
plates. Lead plates were doped by 3\% of antimony to improve their
rigidity. Scintillator plates were made of polystyrene doped by 1.5\%
of paraterphenile and 0.04\% of POPOP. Scintillator was manufactured
at the scintillator workshop of IHEP with the use of molding
technology. The physical properties of the modules are presented in
Table~\ref{tab:module}.
\begin{table}[htp]
  \begin{center}
    \begin{tabular}{ll}\hline
      lead plate thickness              & 0.275~mm \\
      scintillator plate thickness      & 1.5~mm \\
      number of layers                  & 380 \\
      effective radiation length, $X_0$ & 34~mm \\
      total radiation length            & $20 X_0$ \\
      effective Moliere radius, $R_M$   & 59~mm \\
      module size                       & $110\times 110\times
                                          675~\mbox{mm}^3$ \\
      module weight                     & 18~kg \\ \hline
    \end{tabular}
    \caption{Physical properties of the module.}
    \label{tab:module}
  \end{center}
\end{table}
The WLS optical fibers BCF-91A with a diameter of 1.2~mm were used in
the modules. Each fiber penetrated through the module along its
longitudinal axis twice, forming a loop at the face end of the
module. Radius of the loop was 28~mm. In total 72 such looped
fibers formed a grid of $12\times 12$~fibers per module with
spacing of 9.3~mm. All 144 fiber ends were assembled into a bundle of
a diameter about 10 mm, glued, cut, polished and attached to the
photodetector at the downstream end of the module. No optical grease
was used to provide an optical contact between the bundle cap and the
photodetector, thus there was a natural air gap between them.  A
photomutiplier Hamamatsu R5800 was used as a photodetector for the
prototype. The diameter of the photocathode is 25.4~mm, the number of
dynodes is 10, the applied high voltage was about 1100~V.  Each
photomultiplier was monitored by LED light guided to the
photocathodes by a clear polystyrene fiber.


\section{Experimental setup for the prototype studies}
\label{sec:Setup}

\begin{figure}[ht]
  \centerline{
    \includegraphics[width=0.7\hsize]{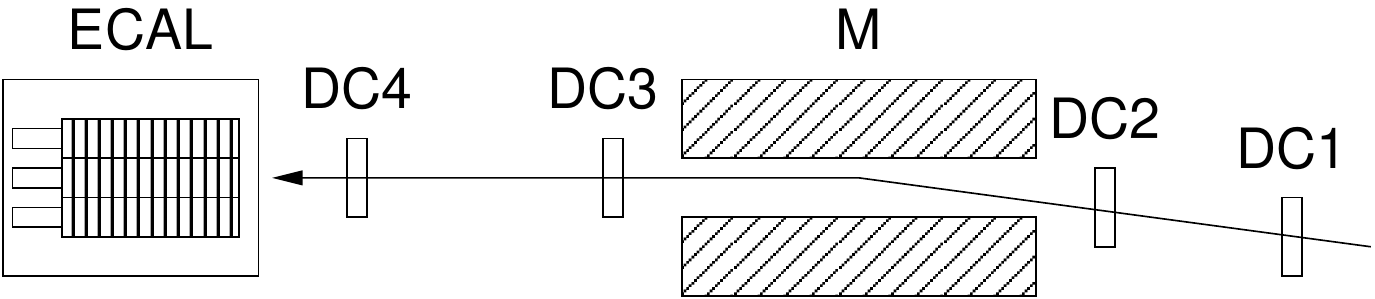}
  }
  \caption{Beam tagging system for the calorimeter prototype studies.}
  \label{fig:channel-2B}
\end{figure}

The prototype of electromagnetic calorimeter consisted of 9 modules
assembled into $3 \times 3$~matrix installed on the remotely
controlled $x,y$-moving support positioned the prototype across the
beam with a precision of 0.4~mm.  The beam line 2B of the U-70
accelerator was used to study performance of the calorimeter
prototype. The secondary beam of negatively charged particles of
momenta from 1 to 19~GeV/$c$ contained more than 70\% of electrons
mixed with muons and hadrons (mainly $\pi^-$ and $K^-$).  Particle
identification was not available at this beam line. A momentum spread
of the beam was at the level of 1 to 5\% at energies from 45 to 1 GeV,
respectively. However, the momentum tagging
system\cite{channel2B-tagging} gave a beam momentum resolution from
0.13\% to 2\% in the same momentum range. The tagging system
illustrated in Fig.\ref{fig:channel-2B} consisted of the dipole magnet
{\tt M} and 4 sets of 2-coordinate drift chambers {\tt DC1}--{\tt
DC4}. A bending angle of the magnet was 55~mrad.
A trigger of the experimental setup used the coincidence of
scintillator counters {\tt S1}, {\tt S2} and {\tt S3} installed
upstream before the first drift chamber {\tt DC1}, and a
scintillator counter {\tt S4} installed after the last drift chamber
{\tt DC4} in front of the calorimeter prototype {\tt ECAL}.

An amplitude out of each prototype cell was measured by the 15-bit
charge sensitive ADC modules LRS2285 over 150-ns gate with a
sensitivity of 30 fC/count. To read out time information from the
drift chamber stations TDC, the LRS3377 CAMAC modules were used. Data
acquisition system included a couple of crates with ADC and TDC
modules as well as control modules to synchronize a read-out
process. VME crate with CAMAC parallel branch driver and PCI-VME
bridge linked all the electronics into the complete system.  Detailed
description of the data acquisition system and front-end electronics
can be found in \cite{btev-nim2}.


\section{Monte Carlo simulations}
\label{sec:MonteCarlo}

The relevant simulation tools were developed. These tools, at the
first stage, are intended mainly for cross-check of experimental
results as well as for tuning of the reconstruction algorithms.

Having proved the consistency of Monte Carlo and the real data, we
plan to use these tools for further optimization of module design and
reconstruction algorithms to provide better performance of the photons
and $\pi^0$'s reconstruction.  Simulation studies were performed with
GEANT3 as a Monte Carlo engine with detailed description of materials
and module geometry.

The developing shower produces light which originates from two
different sources:
\begin{itemize}
\item scintillation in plastic plates due to continuous energy losses 
  when charged particles pass through the active calorimeter material,
\item Cherenkov radiation when charged particles pass through the WLS
  fibers.
\end{itemize}
The simplified technique consists of counting energy deposition in the
active material (with some corrections to take into account light
attenuation in the fibers) and ignoring Cherenkov radiation inside the
fibers. This method is very fast while can not reproduce all details
of the calorimeter response such as non-uniformity due to fibers and
cell borders.
 
For these studies, the detailed light propagation was applied taking
into account the optical properties of the materials, internal
reflections at plate borders, light capture by fibers with double
cladding and the Cherenkov light production and propagation inside the
fibers. It was assumed that attenuation length was 70~cm in the
scintillator and 400~cm in the fiber, scintillator refraction index is
1.59, total internal reflection efficiency at large scintillator faces
is 0.97 and reflection of diffusion type was assumed at side
scintillator faces with the same probability. The mean deposited
energy for one optical photon production in the scintillator was
assumed to be 100~eV and the Cherenkov photons were generated by
GEANT.


\section{Results}
\label{sec:Results}

\subsection{Calibration of the modules}
\label{ssec:calibraton}

\begin{figure}[htb]
  \includegraphics*[width=0.48\hsize]{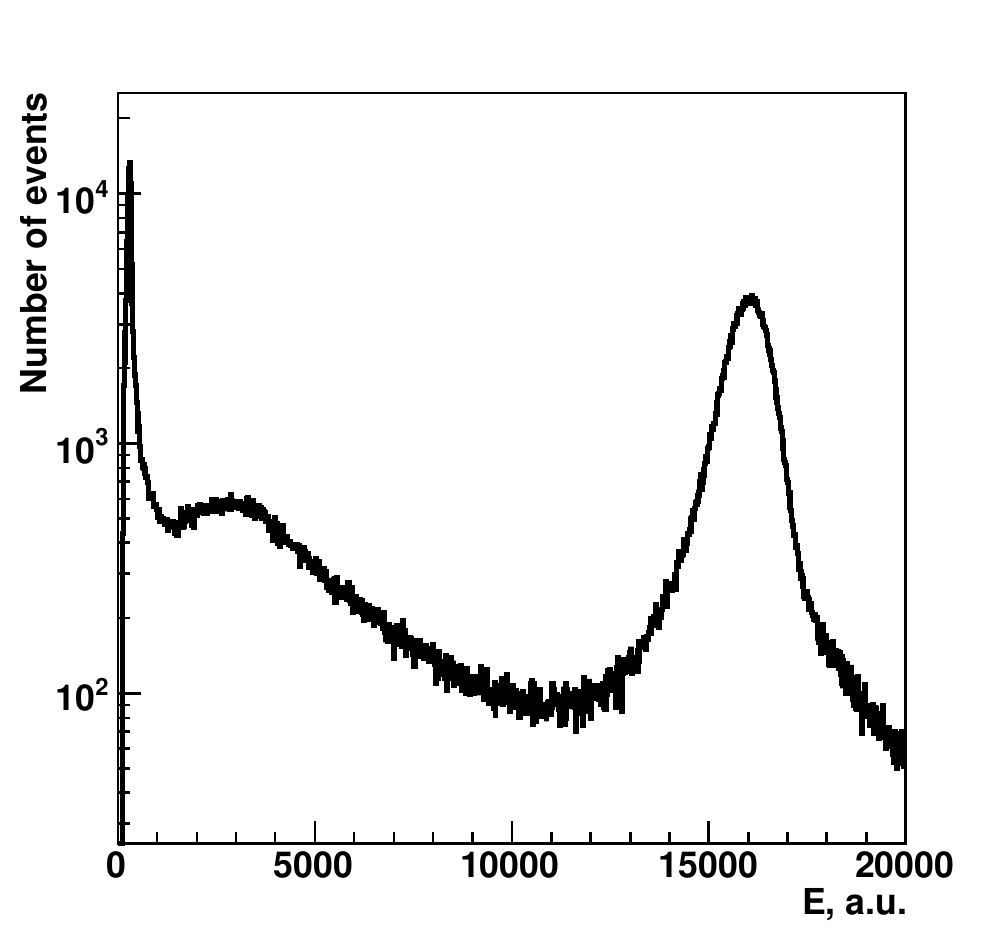}
  \hfill
  \includegraphics*[width=0.48\hsize]{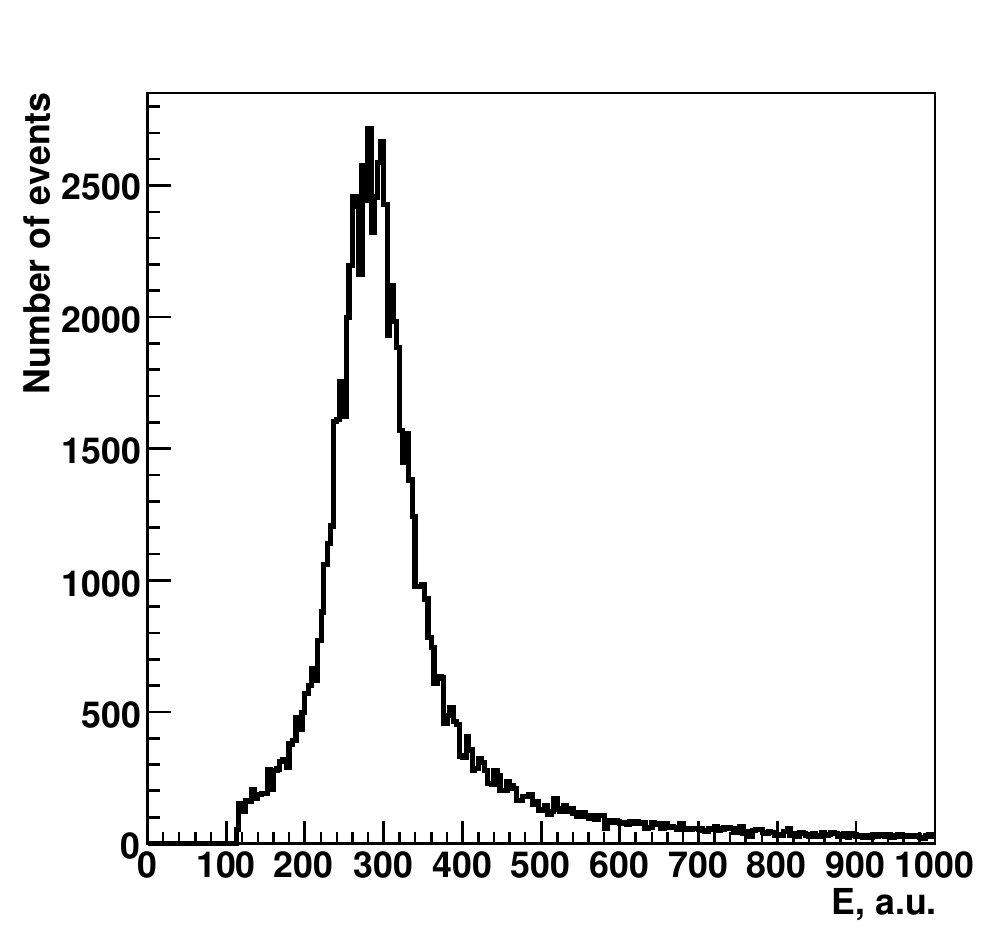}
  \caption{Energy deposited by the 19-GeV/c beam in one module.}
  \label{fig:ECAL-spectrum}
\end{figure}

The modules were calibrated by a 19-GeV/$c$ beam. Each module was
exposed to the beam using an $x,y$-moving support. The energy spectrum
from one module (Fig.\ref{fig:ECAL-spectrum}, left plot)
shows a peak at 19~GeV corresponding to the energy deposited by
electrons. Another peak at low energies is due to minimum ionizing
particles (MIP). A broad distribution in the energy specrtum between
the two peaks is due to hadrons. Calibration of the modules was
possible using both electron and MIP signals, but the best relative
calibration coefficients were found by equalizing MIP signals, while
the absolute calibration was obtained by setting the total measured
energy in the $3\times 3$ matrix to 19~GeV. Events when only one
module has an energy above the threshold of 100~MeV were selected for
the MIP calibration. The energy distribution around the MIP peak
(Fig.\ref{fig:ECAL-spectrum}, right plot) has two contributions. One
is caused by the Landau distribution of ionization energy loss,
and another one is due to the finite energy resolution of the
calorimeter at low energy. The MIP peak was fitted by the Gaussian,
and the mean value of the fitting function served for the relative
calibration.

\subsection{Energy and position resolution}
\label{ssec:E,x resolution}

\begin{figure}[htb]
  \centering
  \includegraphics*[width=0.60\hsize]{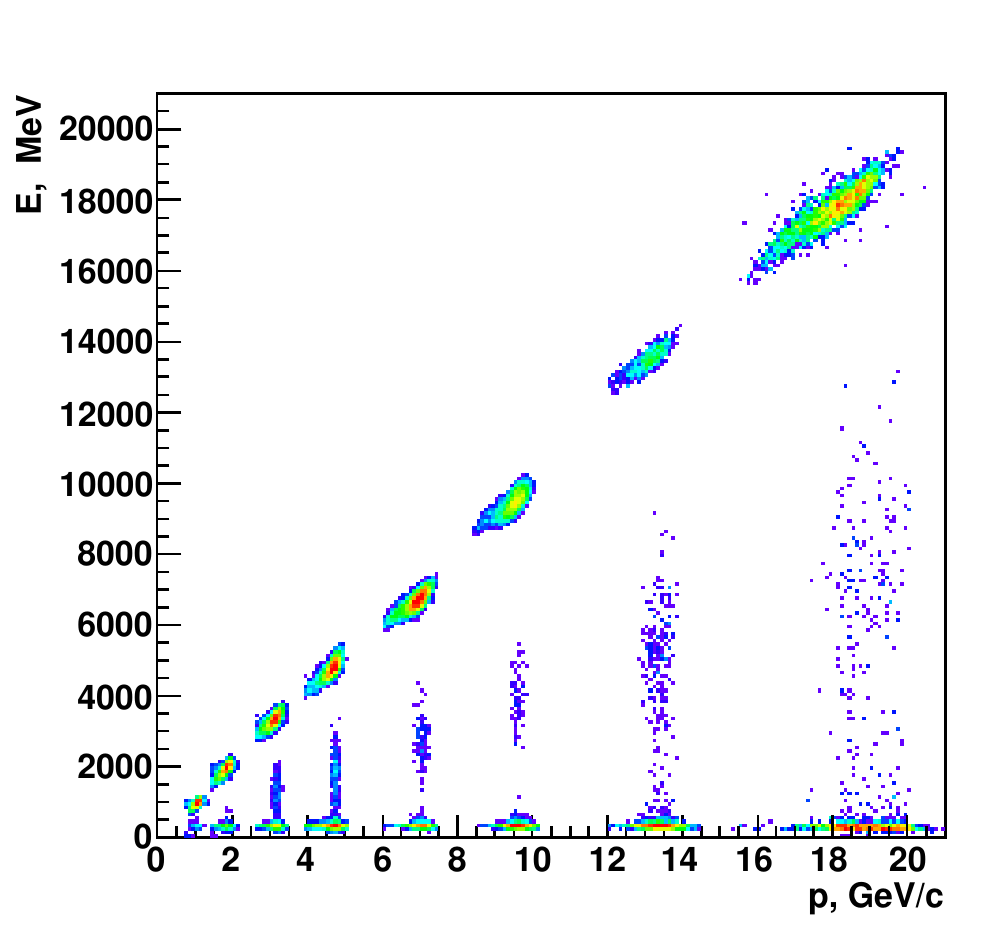}
  \caption{Correlation between the energy measured in the calorimeter
    and the beam momentum measured in the magnetic spectrometer.}
  \label{fig:E vs P}
\end{figure}

\begin{figure}[htb]
  \centering
  \includegraphics*[width=0.60\hsize]{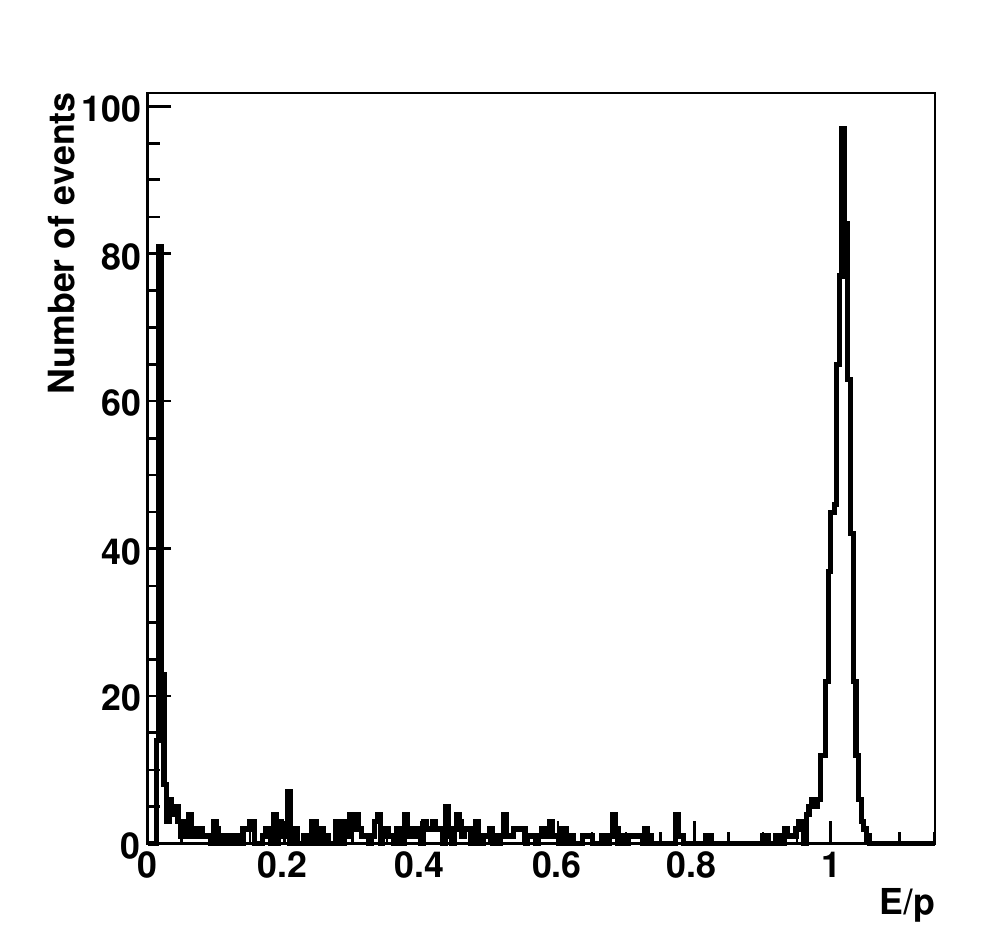}
  \caption{Ratio of the energy $E$ measured in the calorimeter to the
    momentum $p$ measured by the magnetic spectrometer at 19~GeV/$c$.}
  \label{fig:EoverP_19}
\end{figure}

\begin{figure}[htb]
  \centering
  \includegraphics*[width=0.60\hsize]{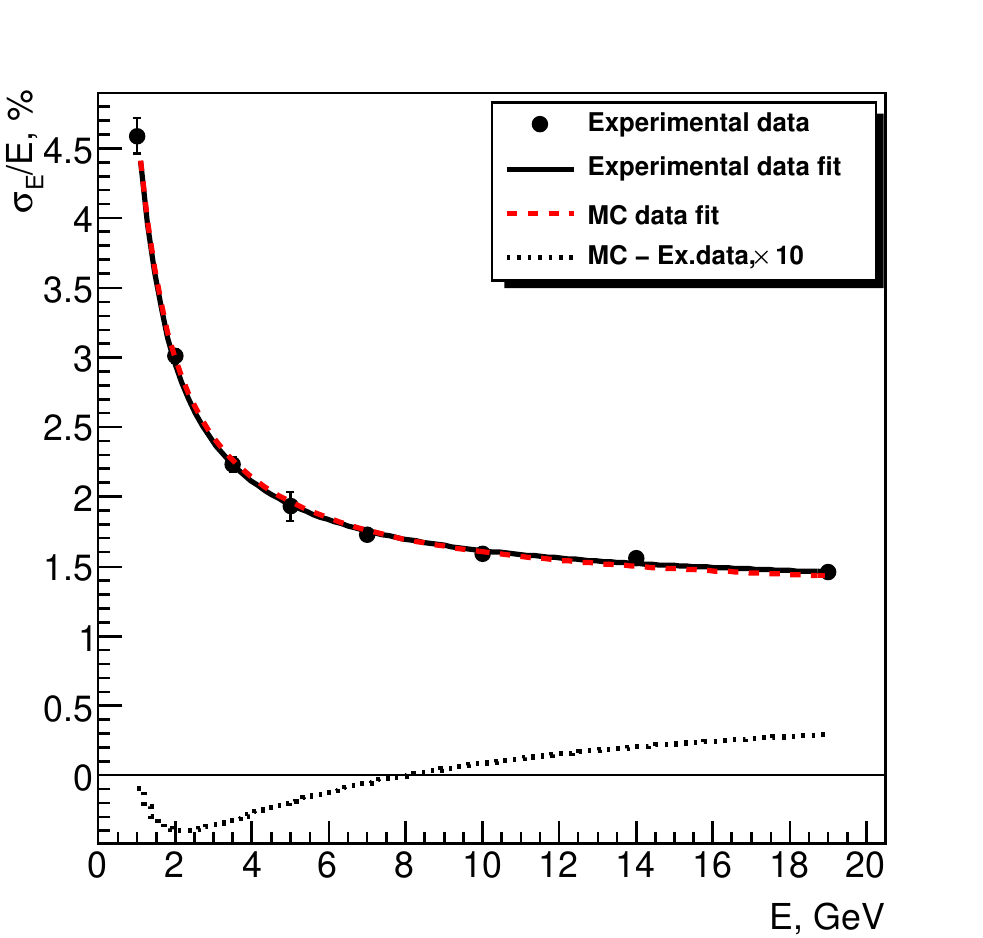}
  \caption{Measured energy resolution.}
  \label{fig:ECAL-eresolu}
\end{figure}

After some dedicated calibration runs, when each module was
exposed to the 19-GeV/$c$ beam, the ECAL prototype was fixed so that
the beam hit the central module. It was exposed to beam at
momenta 1, 2, 3.5, 5, 7, 10, 14 and 19~GeV/$c$. For each beam
momentum, magnetic field in the spectrometric magnet {\tt M} was
adjusted to provide the same bending angle of the beam. The momentum
of the beam particle $p$ was measured by the magnetic spectrometer, and
the energy $E$ measured in the calorimeter prototype is linearly
correlated with the momentum $p$, as illustrated by Fig.\ref{fig:E vs P}.
Therefore, in order to obtain a true energy resolution, the measured
energy should be corrected by the beam momentum, or the energy
resolution can be represented by the width of the distribution of the
$E/p$ ratio (Fig.\ref{fig:EoverP_19}).
The energy resolution is obtained from the Gaussian fit of the right
peak around $E/p=1$. The energy resolution $\Delta E/E$ measured by
electrons at energies from 1 to 19~GeV are shown in
Fig.\ref{fig:ECAL-eresolu}.
The black bullets represent the experimentally measured points. The
solid curve is a result of a fit of these experimental points, and the
dashed curve is a result of a fit of the Monte Carlo points. The
fitting function can be represented by the equation (\ref{eq:efit}):
\begin{equation}
  \frac{\Delta E}{E} =
  \sqrt{\left(\frac{a}{E}\right)^2 +
    \frac{b^2}{E} + c^2},
  \label{eq:efit}
\end{equation}
where parameters $a$, $b$ and $c$ for the experimental and Monte Carlo
fits are shown in Table~\ref{tab:eresolu}.
\begin{table}[ht]
  \centering
  \begin{tabular}{|l|c|c|c|}\hline
    ~ & $a$, $10^{-2}$\,GeV & $b$, $10^{-2}$\,GeV$^{1/2}$ & $c$, $10^{-2}$\, \\ \hline
    Experimental fit & $3.51 \pm 0.28$ & $2.83 \pm 0.22$ & $1.30 \pm 0.04$ \\
    Monte Carlo fit  & $3.33 \pm 0.12$ & $3.07 \pm 0.08$ & $1.24 \pm 0.02$ \\
    \hline
  \end{tabular}
  \caption{Fitting function parameters for the energy resolution.}
  \label{tab:eresolu}
\end{table}
A linear term $a$ of the energy resolution expansion is determined by
a beam spread rather than the calorimeter properties. As it was shown
in previous studies performed at this 2B beam channel
\cite{btev-nim2}, the main contribution to this term comes from the
electronics noise and the multiple scattering of the beam particles on
the beam pipe flanges and the drift chambers. The beam momentum spread
was introduced into Monte-Carlo simulations in order to fully
reproduce the experimental conditions. A simulated energy resolution
is shown by the red dashed line in Fig.\ref{fig:ECAL-eresolu}. The
dotted line at this plot is a difference between the experimental data
fit and the Monte Carlo fit multiplied by 10. Thus a deviation of the
experimental result from the simulation one is less than 0.04\%. The
energy resolution obtained in Monte Carlo is in a good agreement with
the experimental data.

\begin{figure}[htb]
  \centering
  \includegraphics*[width=0.60\hsize]{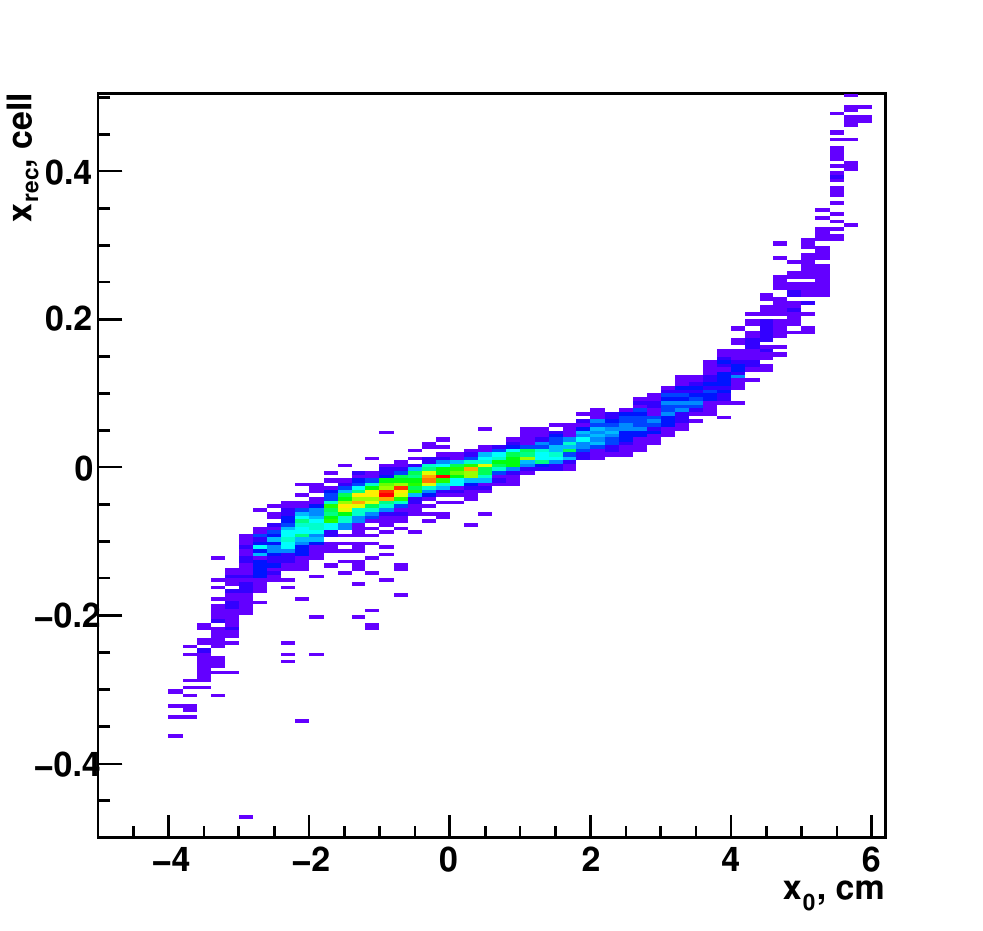}
  \caption{Center of gravity of the electromagnetic shower $X_{\rm
  rec}$ vs the impact coordinate of the electron $x_0$.}
  \label{fig:S-curve}
\end{figure}

\begin{figure}[htb]
  \centering
  \includegraphics*[width=0.60\hsize]{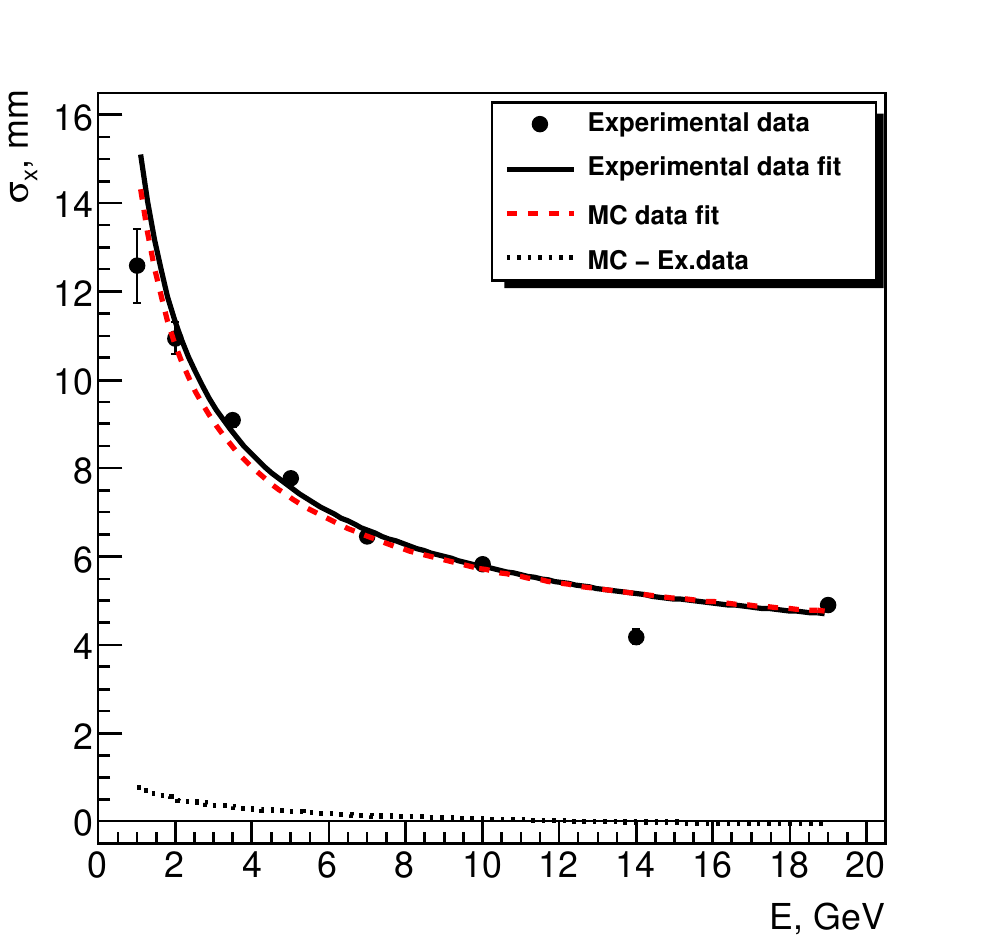}
  \caption{Measured position resolution.}
  \label{fig:ECAL-xresolu}
\end{figure}

Position resolution has been determined by a comparison of the exact
impact coordinate of the beam particle, measured by the last drift
chamber {\tt DC4}, and the center-of-gravity of electromagnetic
shower developed in the calorimeter prototype. Fig.\ref{fig:S-curve}
shows a dependence of the measured coordinate $x_{\rm rec}$ on the
true one $x_0$.
A position resolution in the middle of the module is shown in
Fig.\ref{fig:ECAL-xresolu}, where the bullets represent the
experimentally measured points, the solid curve is a result of the
experimental points fit, and and the red dashed curve is a result of
Monte Carlo points fit.
The data were fitted by the function (\ref{eq:xfit})
\begin{equation}
  \Delta x =
  \sqrt{a^2 + \frac{b^2}{E}},
  \label{eq:xfit}
\end{equation}
where parameters $a$ and $b$ are given in Table~\ref{tab:xresolu}.
\begin{table}[ht]
  \centering
  \begin{tabular}{|l|c|c|}\hline
    ~ & $a$, mm & $b$, mm\,GeV$^{1/2}$ \\ \hline
    Experimental fit & $3.09 \pm 0.16$ & $15.4 \pm 0.3$ \\
    Monte Carlo fit  & $3.40 \pm 0.14$ & $14.5 \pm 0.3$ \\
    \hline
  \end{tabular}
  \caption{Fitting function parameters for the position resolution.}
  \label{tab:xresolu}
\end{table}
The dotted curve in Fig.\ref{fig:ECAL-xresolu} stands for a
deviation of the experimental data fit results from the Monte Carlo
fit results, which are consistent within 5\% of precision.

\subsection{Lateral non-uniformity}
\label{ssec:non-uniformity}

\begin{figure}[htb]
  \centering
  \includegraphics*[width=0.60\hsize]{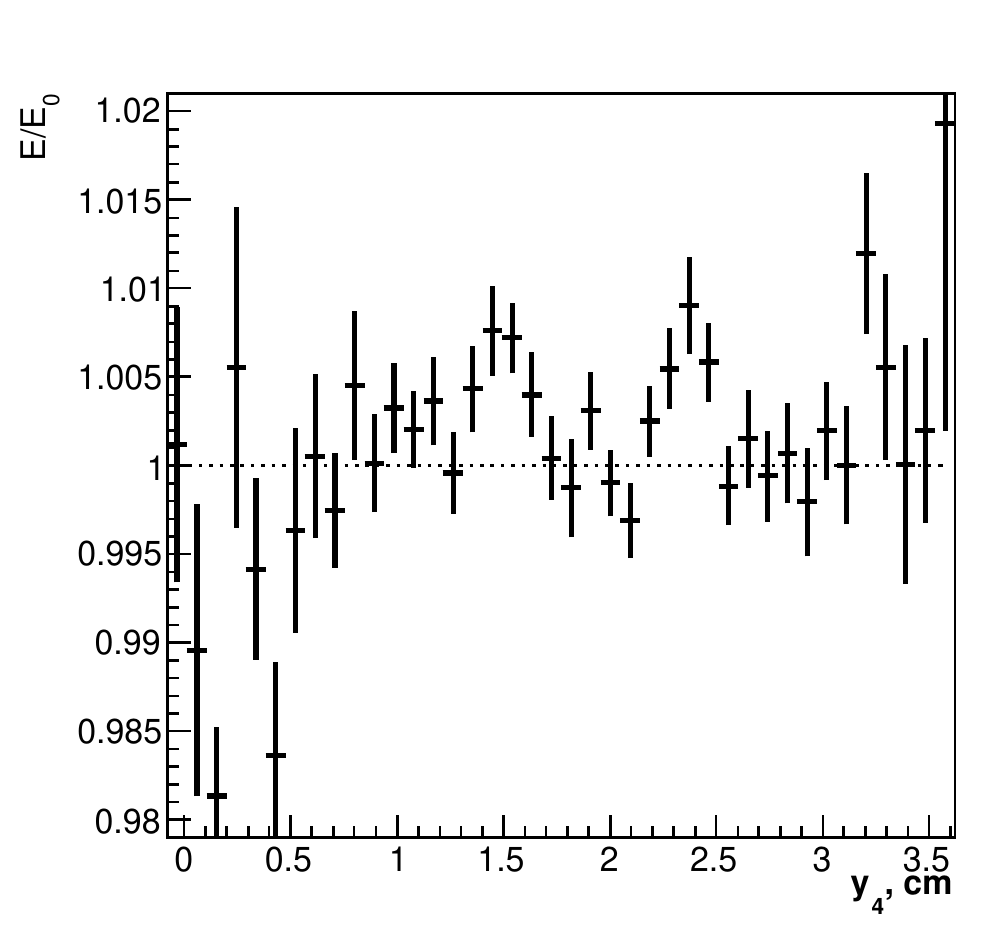}
  \caption{Relative energy response profile vs $y$-coordinate at fixed $x$.}
  \label{fig:EvsY}
\end{figure}

Due to various mechanical inhomogeneities of the prototype one can
expect to observe the dependence of the energy $E$ deposited in the
calorimeter on the hit coordinates $(x,y)$. The ``hot'' zones, if any,
should be seen at the WLS fiber positions, at the steel strings, and
at the boundaries between the modules. A possible lateral
non-uniformity of the energy response was studied with the data
collected in the 19-GeV/$c$-run. The last drift chamber {\tt DC4} was
used to measure the coordinate of the beam particle incidence onto the
calorimeter surface. As the beam contained several particle species
which interact differently with the calorimeter medium (see
Fig.\ref{fig:ECAL-spectrum}), the mean deposited energy was measured
as a function of $(x,y$) for two energy intervals, $E<0.5$~GeV and
$16<E<22$~GeV corresponding to the MIP peak and that of the
electromagnetic shower, respectively. The relative energy response
profile for electrons vs $y$-coordinate at fixed $x$ is shown in
Fig.\ref{fig:EvsY}.
As one can see, the fluctuations of the energy response do not exceed
1\%, that is no lateral non-uniformity of the energy response is
observed within the available statistics.

\subsection{Light output measurement}
\label{ssec:Light yield}

\begin{figure}[htb]
  \includegraphics*[width=0.48\hsize]{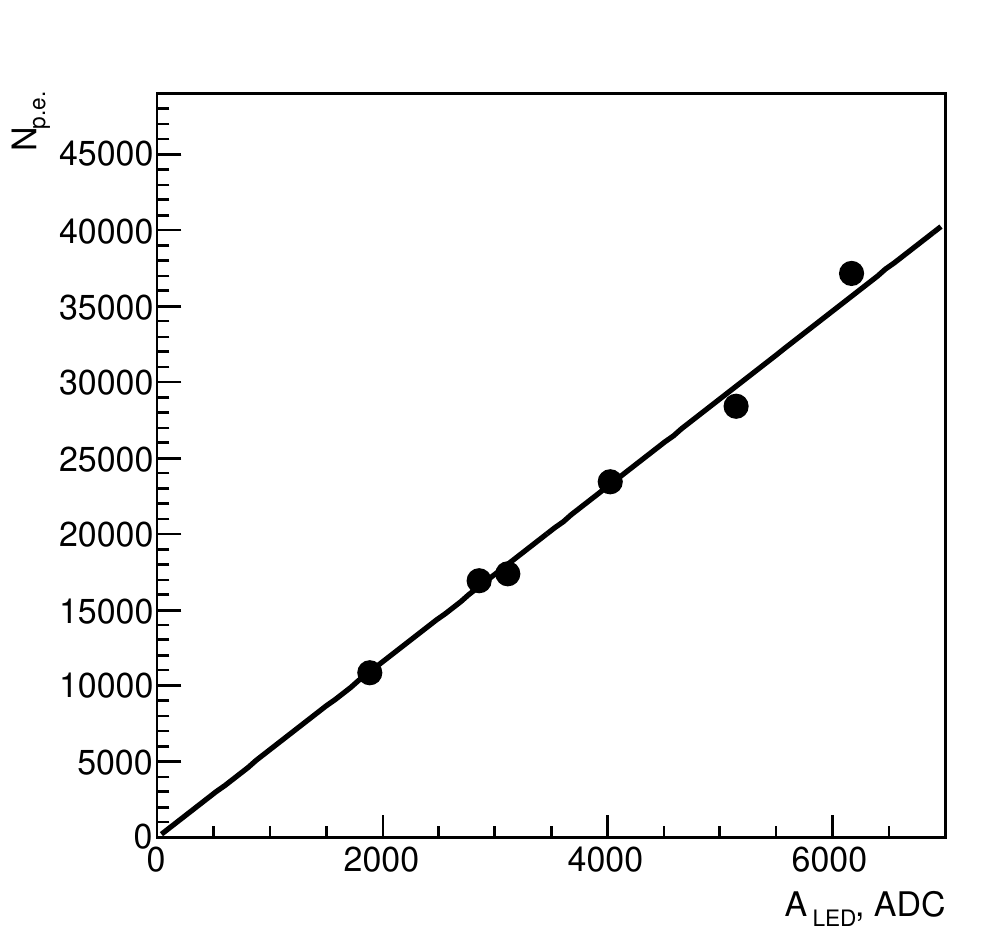}
  \hfil
  \includegraphics*[width=0.48\hsize]{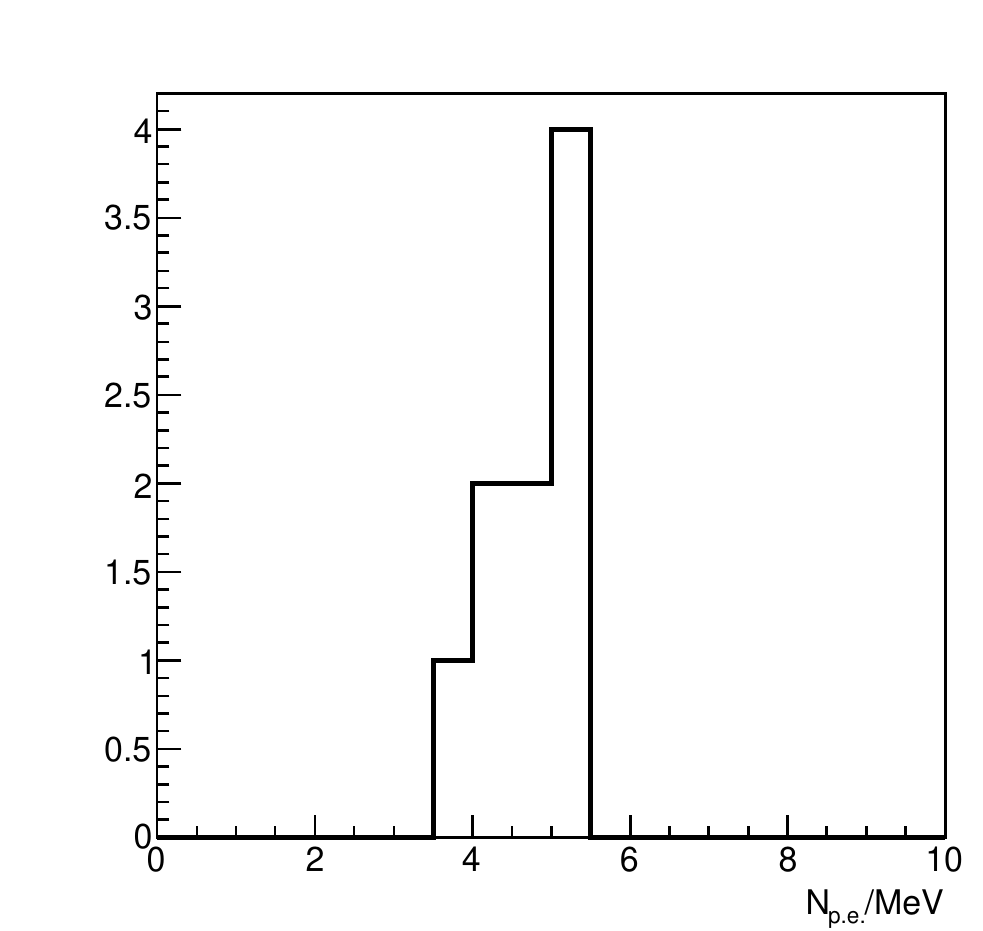}
  \caption{Light output $N_{\rm p.e.}$ vs the LED amplitude for 1
    modules (left) and the light output distribution of all 9 modules
    (right).}
  \label{fig:photostat}
\end{figure}

Light output of the prototype modules, expressed as a number of
photoelectrons $N_{\rm p.e.}$, was evaluated with the highly stable
LED pulses \cite{btev-nim4}. Fluctuations of the measured amplitude $A$
is determined by statistical fluctuations of the number of
detected photoelectrons and by fluctuations of the photomultiplier
gain $M$ \cite{Kowalski1970}:
\begin{equation}
  \left(\frac{\sigma_A}{A}\right)^2 =
  \frac{1}{N_{\rm p.e.}} \left[1+\left(\frac{\sigma_M}{M}\right)^2\right].
  \label{eq:amp_fluct}
\end{equation}
The gain fluctuation can be defined through the secondary emission
factor of the first dynode $\delta_1$ and the secondary emission
factor of other dynodes $\delta$:
\begin{equation}
  \left(\frac{\sigma_M}{M}\right)^2 =
  \frac{\delta}{\delta_1} \cdot \frac{1}{\delta-1}.
  \label{eq:gain_fluct}
\end{equation}
The total gain of the 10-dynode photomultiplier R5800 was equal to
$10^6$ for the applied high voltage 1100~V, and the potential of the
first dynode was boosted to increase the secondary emission factor
$\delta_1$ to approximately 5. Thus, one can
obtain the emission factor $\delta=3.9$ and the equation
(\ref{eq:amp_fluct}) is derived to the number of the detected
photoelectrons as a function of the relative amplitude width:
\begin{equation}
  N_{\rm p.e.} \approx \frac{1.3}{(\sigma_A/A)^2}.
  \label{eq:photostat}
\end{equation}
A set of runs with six different LED amplitudes has been carries
out. A dependence of the number of photoelectrons on the LED
amplitude for one cell and a distribution of the light output for all
9 cells are shown in Fig.\ref{fig:photostat}.
These plots were fitted by the linear function, which slope represents
the number of photoelectrons per one ADC count. Being divided by the
calibration coefficient, one can obtain that the number of
photoelectrons detected by the the prototype modules is $4.8 \pm
0.6$~p.e./MeV.


\section*{Conclusion}
\label{sec:Conclusion}

The measurements of energy and position resolutions of the
electromagnetic calorimeter prototype of fine-sampling type for the
PANDA and CBM experiments at FAIR at Darmstadt have been carried out
at the IHEP test beam facility at the Protvino 70~GeV accelerator.
The prototype consisted of a $3\times 3$~array with the cell sizes of
$11 \times 11$~cm$^2$.  Each cell had 380 layers with 1.5~mm
scintillator and 0.3~mm lead.  Scintillation light was collected
by optical fibers penetrating through the modules longitudinally along
the beam direction. The prototype was designed and assembled at the IHEP
scintillator workshop.

Studies were made in the electron beam energy range from 1 to
19~GeV. The energy tagged has allowed us to measure the stochastic
term in energy resolution as $(2.8 \pm 0.2)\times
10^{-2}~\mbox{GeV}^{1/2}$ which is consistent with the one measured
at BNL for the KOPIO project in the energy range from 0.05~GeV to
1~GeV.  
Taking into account the effect of light transmission in
scintillator tiles and WLS fibers, photo statistics as well as noise of
the entire electronic chain resulted in good agreement between the
measured energy resolutions and the GEANT Monte Carlo simulations.

The stochastic term in the dependence of position resolution on energy
in our measurements is about $15.4 \pm 0.3$~mm\,GeV$^{1/2}$ which is
in agreement with Monte Carlo simulations. 
For 10~GeV electrons
position resolution is 6~mm in the center of the cell, and is 3~mm at
a boundary between two cells.

The non-uniformity of the energy response of the prototype due to
holes for straight fibers studied with the use of electrons and MIPs
has turned out to be negligible. Monte-Carlo simulations are in a good
agreement with the obtained experimental results.

The characteristics experimentally determined for our calorimeter
prototype well meet the design goals of the PANDA and CBM
experiments. However, the final conclusion on lateral sizes of the
cells as well as on Shashlyk longitudinal sampling structure could be
done only after studies of reconstruction efficiency of $\pi^0$-mesons
of different energies. 


\section*{Acknowledgment}
\label{sec:Acknowledgment}

This work was partially supported by the Rosatom grant with
Ref. No. \linebreak N.4d.47.03.08.118, by the INTAS grants with Ref. No.
06-1000012-8845 and 06-1000012-8914, and by the RFBR grant
05-02-08009. 




\end{document}